\documentclass[reprint,superscriptaddress, preprintnumbers, nofootinbib,amsmath,amssymb,aps]{revtex4-2}
\pdfoutput=1

\usepackage{graphicx}% Include figure files
\usepackage{dcolumn}% Align table columns on decimal point
\usepackage{bm}% bold math
\usepackage{caption}
\usepackage{subcaption}
\usepackage{amsmath}
\usepackage{hyperref}
\usepackage[utf8x]{inputenc}
\hypersetup{colorlinks=true,allcolors=blue}

\begin{document}

\preprint{MITHIG-TH-22-001, CERN-TH-2022-057}

\title{Data-driven extraction of the substructure of quark and gluon jets in proton-proton and heavy-ion collisions}

\author{Yueyang Ying}
\affiliation{Relativistic Heavy Ion Group, MIT, Cambridge, Massachusetts, USA}
\author{Jasmine Brewer}
\affiliation{Theoretical Physics Department, CERN, CH-1211 Gen\`eve 23, Switzerland}
\author{Yi Chen}
\affiliation{Relativistic Heavy Ion Group, MIT, Cambridge, Massachusetts, USA}
\author{Yen-Jie Lee}
\affiliation{Relativistic Heavy Ion Group, MIT, Cambridge, Massachusetts, USA}

\begin{abstract}
The different modifications of quark- and  gluon-initiated jets in the quark-gluon plasma (QGP) produced in heavy-ion collisions is a long-standing question that has not yet received a definitive answer from experiments. In particular, the relative sizes of the modification of quark and gluon jets differ between theoretical models. Therefore, a fully data-driven technique is crucial for an unbiased extraction of the quark and gluon jet spectra and substructure. We perform a proof-of-concept study based on proton-proton and heavy-ion collision events from the \textsc{Pyquen} generator with statistics accessible in Run 4 of the Large Hadron Collider. We use a statistical technique called topic modeling to separate quark and gluon contributions to jet observables. We demonstrate that jet substructure observables, such as the jet shape and jet fragmentation function, can be extracted using this data-driven method. These values can then be used to obtain the modification of quark and gluon jet substructures in the QGP. We also perform the topic separation on smeared input data to demonstrate that the approach is robust to fluctuations arising from a QGP background. These results suggest the potential for an experimental determination of quark and gluon jet spectra and their substructure.
\end{abstract}

\maketitle

\noindent
During the first millionth of a second after the Big Bang, the universe comprised hot and dense primordial matter of deconfined quarks and gluons before cooling down and forming ordinary matter. This deconfined phase of matter, the quark-gluon plasma (QGP), only exists at extremely high temperatures and pressures and is recreated on earth in high-energy heavy-ion collisions (see~\cite{big-picture} for a review).

High-energy collisions between protons or nuclei occasionally produce very high-energy quarks or gluons that successively fragment and hadronize into collimated sprays of particles called jets. Jets are a ubiquitous tool for studying Quantum Chromodynamics (QCD) and have been widely used in the studies of both proton-proton~\cite{Kogler:2018hem} and heavy-ion collisions~\cite{Qin:2015srf,Connors:2017ptx,Cunqueiro:2021wls,Apolinario:2022vzg} at the Large Hadron Collider (LHC), the Relativistic Heavy Ion Collider (RHIC), and recently in electron-positron annihilation~\cite{Chen:2021uws} with archived ALEPH data at the Large Electron-Position Collider~\cite{Badea:2019vey}.

The hot quark-gluon plasma produced in heavy-ion collisions modifies the properties of jets. High-energy partons propagating through the QGP lose energy due to multiple elastic scatterings and medium-induced gluon radiation~\cite{Baier:1996sk,Baier:1996kr,Zakharov:1996fv,Gyulassy:2000fs,Gyulassy:2000er,Wiedemann:2000za,Salgado:2003gb,Majumder:2009ge,Guo:2000nz,Wang:2001ifa,Chesler:2014jva,Casalderrey-Solana:2014bpa,Schenke:2009gb,Park:2018acg}, often referred to as jet quenching~\cite{Bjorken:1982tu,ATLAS:2010isq,CMS:2011iwn}. The resulting suppression of the yield of jets in heavy-ion collisions compared to an equivalent number of proton-proton collisions has been observed~\cite{ATLAS:2012tjt,ALICE:2015mjv,CMS:2016uxf,ATLAS:2018gwx,ALICE:2019qyj,CMS:2021vui,STAR:2017hhs}. The structure of jets is also modified~\cite{CMS:2013lhm,CMS:2014jjt,ATLAS:2014dtd,ATLAS:2017nre,CMS:2018mqn,CMS:2018jco,ATLAS:2019dsv} in heavy-ion collisions and is an important tool for studying the properties of the quark-gluon plasma.

Quarks and gluons interact with the QCD medium proportional to their color charge, meaning that gluons interact more with the medium than quarks by a factor $C_A/C_F = 9/4$. However, a jet initiated by a quark or gluon quickly fragments into both quarks and gluons. Understanding the modification of quark- and gluon-initiated jets in the quark-gluon plasma may shed light on how the quark-gluon plasma resolves color structure within a jet (see e.g.~\cite{chien2018probing,Mehtar-Tani:2018zba,Qiu:2019sfj,Apolinario:2020nyw}). 

Experimentally, distinguishing quark and gluon jets is challenging since jet measurements are a combination of jets initiated by both. There has been extensive work on data-driven techniques for distinguishing quark and gluon jets in proton-proton~\cite{metodiev-topic-of-jets,komiske,light-quark,Jones:1988ay,Fodor:1989ir,qg-sub,qg-tagging,Gras:2017jty,Frye:2017yrw,Larkoski:2019nwj,Dreyer:2021hhr} and heavy-ion collisions~\cite{chien2018probing,brewer,jet-charge}. Especially in heavy-ion collisions with substantial theoretical uncertainties in Monte Carlo event generators, it is highly advantageous to distinguish quark and gluon jets in a way that does not rely on Monte Carlo labeling. 
In this case, we wish to identify two physics-motivated categories (quark- and gluon-initiated jets) underlying unlabeled jet measurements. 
Prior work has demonstrated success in using a statistical technique called topic modeling to distinguish quark- and gluon-initiated jets in both proton-proton~\cite{metodiev-topic-of-jets,komiske} and heavy-ion~\cite{brewer} collisions, using two measurable jet samples that differ in their quark- and gluon-initiated jet fractions.

In this work, we apply topic modeling to dijet and photon-jet ($\gamma+$jet) samples from \textsc{Pyquen}~\cite{pyquen}, which is a Monte Carlo event generator that simulates medium-induced energy loss of partons in heavy-ion collisions. In Section~\ref{sec:topics} we discuss the topic modeling approach, which relies on the assumption that dijet and $\gamma$+jet samples are mixtures of the same underlying quark- and gluon-initiated jet distributions, except with different quark and gluon fractions. In Section~\ref{sec:simulations} we discuss the samples we use, and in Section~\ref{sec:topicresults} we use jet constituent multiplicity distributions to extract the quark and gluon fractions in these jet samples, separately in simulations of proton-proton and heavy-ion collisions. In Section~\ref{sec:substructure} we use these fractions to extract quark and gluon jet substructure observables in proton-proton and heavy-ion collisions. The quark and gluon jet substructure accessed using this data-driven method agree qualitatively with the substructure of quark- and gluon-initiated jets as defined from Monte Carlo-level information. This suggests the potential for an entirely data-driven procedure to experimentally extract quark and gluon jet substructure and their modification. In Section~\ref{sec:thermal} we show that these results are robust to Gaussian smearing of the multiplicity distribution, suggesting the possibility of using this technique in the large background present in heavy-ion collisions. 

\section{Topic Modeling}
\label{sec:topics}

To distinguish quark- and gluon-initiated jets experimentally, jet samples collected from events at RHIC or the LHC can be thought of as unlabeled mixtures of quark and gluon jets. Distinguishing quark and gluon jets from these unlabelled mixtures falls in the broad category of unsupervised learning, where the goal is to infer some structure or pattern in a set of unlabeled data. Following previous work~\cite{brewer,metodiev-topic-of-jets}, we apply the ``topic modeling'' technique to solve this problem. Topic modeling is traditionally used to discover abstract ``topics'' that occur in a collection of text documents. In the context of jet topics, the two categories (``topics'') underlying jet measurements are quark- and gluon-like jets~\cite{metodiev-topic-of-jets}.

To illustrate the concept of topic modeling, we present an example which we will refer to throughout the section. Suppose we have two input distributions, input A (pp $\gamma+$jet sample), and input B (pp dijet sample), as shown below in Fig. \ref{fig:example-input}. In the case of quark/gluon topic modeling, we assume that these two input distributions are both a combination of the same two unknown base distributions, or ``topics'' (one quark-like and one gluon-like). Heuristically, this assumption is based on the intuition that jets can be classified as being initiated by a quark or gluon. These contributions to the example input distributions are shown as dashed curves in Fig.~\ref{fig:example-input}.
Each input sample has a different fraction of each topic. The goal of the algorithm is to derive these underlying distributions from measurements.

\begin{figure}[htp]
    \centering
    \begin{subfigure}{0.23\textwidth}
        \centering
        \includegraphics[width=\textwidth]{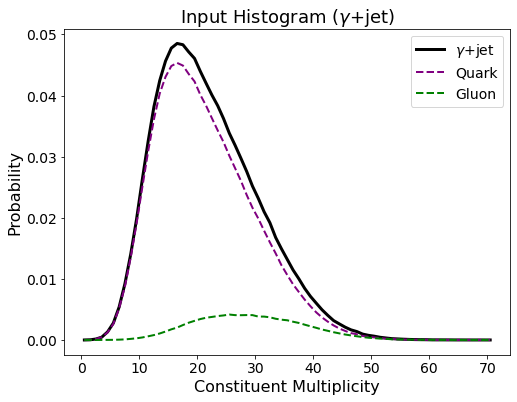}
        \caption{Input A ($\gamma+$jet sample)}
        \label{fig:input-A}
    \end{subfigure}
    \begin{subfigure}{0.23\textwidth}
        \centering
        \includegraphics[width=\textwidth]{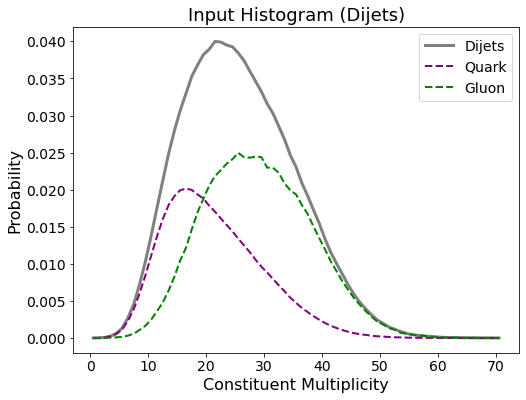}
        \caption{Input B (dijet sample)}
        \label{fig:input-B}
    \end{subfigure}
    \caption{Example of two input distributions that might be used for topic modeling, demonstrating that each input is a superposition of the same underlying base distribution shapes}
    \label{fig:example-input}
\end{figure}

Jet samples collected from colliders are assumed to be mixtures of these jet topics, so that each jet observable histogram is a mixture of the two underlying quark/gluon base distributions. Mathematically, we can then represent the input histograms as
\begin{equation} \label{eq:mix}
    p^{(s)}(x) = f^{(s)} b_1(x) + (1-f^{(s)}) b_2(x)
\end{equation}
where $p^{(s)}(x)$ is the probability density of some observable $x$ in sample $s$, $b_i(x)$ are the base distributions (topics), and $f^{(s)}$ and $1-f^{(s)}$ are the fractions of topic 1 and 2 in sample $s$, respectively. 
However, Eq.~\ref{eq:mix} is ambiguous because there are infinitely many ways to define $b_1$ and $b_2$ and modify $f^{(s)}$ accordingly such that the equation remains true. In order to resolve this, we follow~\cite{metodiev-topic-of-jets} and use the \texttt{DEMIX} algorithm~\cite{katzsamuels2019decontamination}, which breaks this ambiguity by choosing unique base distributions $b_1$ and $b_2$ that satisfy an additional requirement called mutual irreducibility.

\texttt{DEMIX} results in the \textit{mutually irreducible}~\cite{blanchard2016classification} underlying distributions, $b_1$ and $b_2$, that satisfy the requirement that neither contains any contribution from the other. In other words, we cannot write $b_1(x) = c b_2(x) + (1-c)F$, or vice versa, for any probability distribution $F$ and $0 < c \leq 1$. This also implies that $\lim_{x \rightarrow x_\text{max}} b_1(x)/b_2(x) = 0$ and $\lim_{x \rightarrow x_\text{min}} b_2(x)/b_1(x) = 0$, where the probability distributions $b_1, b_2$ are defined on $x \in (x_\text{min},x_\text{max})$ \footnote{Depending on how we define $b_1(x)$ and $b_2(x)$, the limits may be reversed. That is, we may find $\lim_{x \rightarrow x_\text{min}} b_1(x)/b_2(x) = 0$ and $\lim_{x \rightarrow x_\text{max}} b_2(x)/b_1(x) = 0$}.

It is worth noting that \texttt{DEMIX} requires the input distributions to have different purities of the same underlying base distributions and guarantees that the two resulting base distributions are mutually irreducible. Since the Monte Carlo definition of quark and gluon jets is not well-defined, Ref.~\cite{komiske} defines the \textit{operational definition} quark and gluon categories as the mutually irreducible underlying distributions in a jet substructure feature space, given two mixed QCD jet samples at a fixed $p_T$.
For this work, we choose to use constituent multiplicity (number of constituent hadrons in a given jet) as the jet observable because the multiplicities of quark and gluon jets are mutually irreducible in the high-energy limit~\cite{Frye:2017yrw} and it exhibits good performance in proton--proton~\cite{metodiev-topic-of-jets} and heavy-ion~\cite{brewer} studies.

Extracting the base distributions requires finding the reducibility factor $\kappa$, which is the largest amount one distribution that can be subtracted from the other such that all bins remain non-negative,
\begin{equation} \label{kappa}
    \kappa_{ij} = \inf_x \frac{p^{(i)}(x)}{p^{(j)}(x)}\,,
\end{equation}
where $i,j$ index the samples.

\begin{figure}[htp]
    \centering
    \begin{subfigure}{0.23\textwidth}
        \centering
        \includegraphics[width=\textwidth]{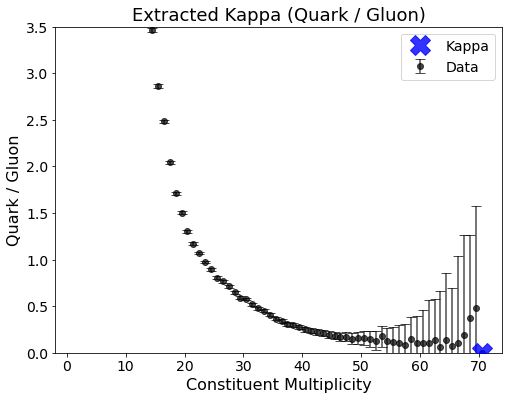}
        \caption{Quark / Gluon (pp)}
        \label{fig:qg}
    \end{subfigure}
    \begin{subfigure}{0.23\textwidth}
        \centering
        \includegraphics[width=\textwidth]{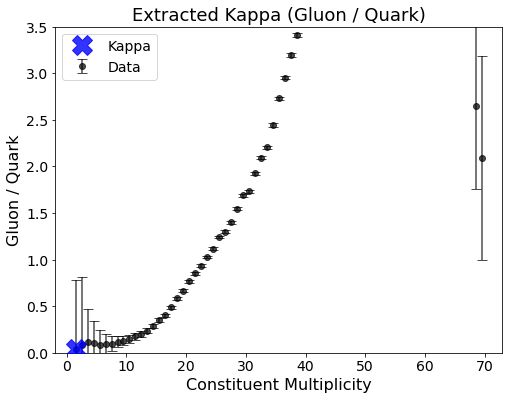}
        \caption{Gluon / Quark (pp)}
        \label{fig:gq}
    \end{subfigure}
    \caption{The ratio of the quark and gluon truth histograms in our example. The extracted $\kappa_{qg}$ and $\kappa_{gq}$ are marked along the tail of the plot, which demonstrates that these base distributions are approximately mutually irreducible.}
    \label{fig:mutually-irred}
\end{figure}

It is worth noting that the mutual irreducibility of the base distributions is the assumption that $\kappa_{qg}$ and $\kappa_{gq}$ are zero. The extracted $\kappa_{qg}$ and $\kappa_{gq}$ are shown in Fig. \ref{fig:mutually-irred}, along with the ratio of each bin in the MC-level quark and gluon jet distributions of our example. The proximity to zero demonstrates the mutual irreducibility of the quark and gluon distributions as defined by the Monte Carlo. 

With the mixture distributions $p_A$ and $p_B$ and the reducibility factors, the base distributions are given by 
\begin{equation} \label{eq:topics}
    \begin{split}
        b_1(x) = \frac{p_A(x) - \kappa_{AB}p_B(x)}{1-\kappa_{AB}},\\
        b_2(x) = \frac{p_B(x) - \kappa_{BA}p_A(x)}{1-\kappa_{BA}}
    \end{split}
\end{equation}

Following Ref.~\cite{brewer}, we extract $\kappa$ by fitting the sampled histograms $p_A$ and $p_B$ to a sum of skew-normal distributions, expressed as
\begin{equation} \label{eq:skew-norm}
    f_N(x; \alpha_i, \theta) = \sum_{k=1}^N \alpha_{i,k} \text{SN}(x; \mu_k, \sigma_k, s_k)
\end{equation}
where $\text{SN}(x; \mu_k, \sigma_k, s_k)$ represents a skew-normal distribution with parameters $\mu_k$, $\sigma_k$, and $s_k$. While the mixture fractions $\alpha_i$ are unique to the input histograms, $ \mu_k, \sigma_k, s_k$ are shared between the two. We use $N=4$, such that we have 18 fit parameters\footnote{The 18 fit parameters are composed of 3 parameters in $\alpha_A$, 3 in $\alpha_B$, and 12 in $\theta$. The fractions represented by $\alpha_A$ and $\alpha_B$, each contain 3 parameters since the last fraction can be calculated: $1-(\alpha_1 + \alpha_2 + \alpha_3)$. Each skew normal distribution is defined by parameters $\mu_k, \sigma_k, s_k$, which gives 12 parameters in $\theta$ when $N=4$.}, represented by $\alpha_A$, $\alpha_B$, and $\theta$~\cite{brewer}.

Assuming that the counts in the histograms follow a Poisson distribution, the best-fit parameters and the corresponding uncertainties can thus be captured by the Poisson-likelihood chi-square function~\cite{BAKER1984437, chi-square, brewer}. In order to extract the parameter values and uncertainties from the likelihood function, we use Markov chain Monte Carlo (MCMC)~\cite{emcee}. We obtain initial estimates of the parameter values by running a simultaneous least-squares fit. For the results shown in this paper, we use 100 MCMC walkers, initialized using the least-squares parameters, and run for 35,000 samples using a burn-in of 30,000 samples.

\begin{figure}[htp]
    \centering
    \begin{subfigure}{0.23\textwidth}
        \centering
        \includegraphics[width=\textwidth]{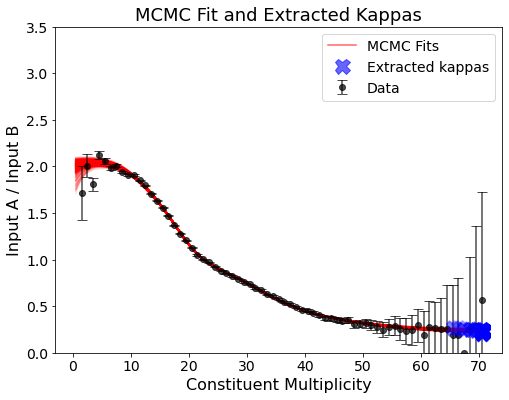}
        \caption{$\gamma+$jet / Dijet (pp)}
        \label{fig:ab}
    \end{subfigure}
    \begin{subfigure}{0.23\textwidth}
        \centering
        \includegraphics[width=\textwidth]{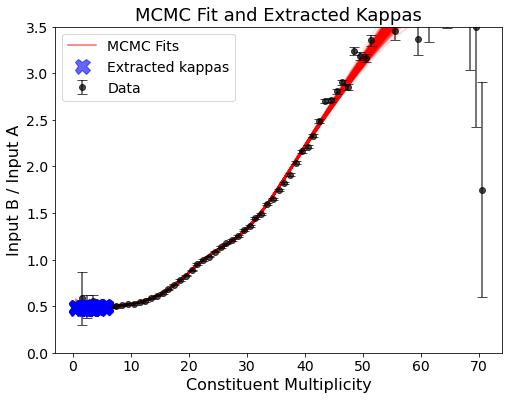}
        \caption{Dijet / $\gamma+$jet (pp)}
        \label{fig:ba}
    \end{subfigure}
    \caption{The ratio of input A and input B histograms, shown along with the MCMC fit and correspondingly sampled $\kappa_{AB}$ and $\kappa_{BA}$ values along the tails.}
    \label{fig:ab-mcmc}
\end{figure}

Once we have the MCMC fits for the input distributions, the reducibility factors in Eq.~\ref{kappa} can be extracted from the ratios of these fits. Fig. \ref{fig:ab-mcmc} displays the ratios of our input distributions, along with the results from the MCMC. Each fit is an element of the posterior distribution of the parameters that is sampled by the MCMC, each with a different minimum, or $\kappa$ value.
In order to extract the topics and calculate the corresponding uncertainty using Eq. \ref{eq:topics}, we sample $\kappa_{AB}$ and $\kappa_{BA}$ from the fits and calculate the mean and standard deviation of the distribution. Here, the reducibility factor for input A/input B, $\kappa_{AB}$, is extracted from the right tail of Fig. \ref{fig:ab}, and similarly, $\kappa_{BA}$ is extracted from the left tail of Fig. \ref{fig:ba}, as that is where the minimum of the curve is located. The sampled reducibility factors are shown in Fig. \ref{fig:ab-mcmc}.

\section{Simulated Collision Events}
\label{sec:simulations}

This proof-of-concept study is based on proton-proton and heavy-ion collision events from the \textsc{Pyquen} generator~\cite{pyquen} with statistics accessible in Run 4 of the Large Hadron Collider. The \textsc{Pyquen} event generator simulates radiative and collisional energy loss of partons in the QGP in heavy-ion collisions~\cite{pyquen}. The input distributions are photon-jet ($\gamma+$jet) and dijet samples. We choose these because at Large Hadron Collider energies, $\gamma+$jet and dijets have different quark and gluon jet fractions. In Section~\ref{sec:substructure}, we consider modified jets in \textsc{Pyquen} that are not embedded in thermal background. We will explore the consequences of thermal smearing in Section~\ref{sec:thermal}.
      
We generate proton-proton (pp) and heavy-ion (PbPb) events at $\sqrt{s}=5.02$~TeV using $\hat{p_T}>80$ GeV, where $\hat{p_T}$ is the hard scattering scale. The impact parameter range for PbPb events corresponds to 0-10\% centrality. We use \textsc{FastJet} 3.3.0~\cite{fastjet,kt} to reconstruct anti-$k_t$ jets with radius $R=0.4$~\cite{antikt}. In the $\gamma+$jet samples, we select the leading jet in the opposite direction to the high-momentum photon ($|\Delta \phi| > \pi/2$), and in the dijet samples, we select the two jets in the event with the largest transverse momenta.

We only include jets with $80 < p_T < 100$ GeV and we impose cuts of $|\eta_\text{jet}| < 1$ and $|\eta_\gamma|<1.442$. 
We additionally remove any jets for which a photon carries more than 80\% of the jet $p_T$.
This removes a low multiplicity peak in the $\gamma+$jet sample that is due to the clustering algorithm incorrectly recognizing a photon with high energy as the leading jet in the opposite azimuthal direction (compared to the high-momentum photon) \footnote{This can be interpreted as a back-to-back photon event, where a photon (plus noise) is incorrectly deemed a jet.}. 

Throughout this work, we will assess the performance of the topic modeling algorithm by presenting comparisons to distributions of quark- and gluon-initiated jets as defined at the Monte Carlo (MC) level.
These labels are only defined at leading order and are therefore not strictly well-defined, but are nonetheless a useful proxy. To determine such labels, we compare the angular distance between the selected jet and the two outgoing matrix elements in the simulation. For $\gamma+$jet, we simply label the jet by the outgoing matrix element that is not the photon. For dijets, we match the jet to the outgoing matrix element with the smallest angular distance $\Delta R = \sqrt{(\Delta \eta)^2 + (\Delta \phi)^2}$ from the jet.
In all MC labels, we only include jets for which $\Delta R < 0.4$ between the matrix element and the jet axis. In pp, 97\% of the jets in the $\gamma$+jet sample and 92\% of those in the dijet sample satisfy the criterion to be given a quark/gluon truth label. In the PbPb sample, we utilize 95\% and 89\% for $\gamma+$jet and dijet quark/gluon truth labels, respectively. Ultimately, the MC labeling in any of the results should not be taken as the absolute truth, but rather as an approximation. Unless otherwise stated, we use the $\gamma$+jet sample for MC quark and gluon labels.

\section{Topic Modeling Results}
\label{sec:topicresults}
\begin{figure}[htp]
    \centering
    \begin{subfigure}{0.23\textwidth}
        \centering
        \includegraphics[width=\textwidth]{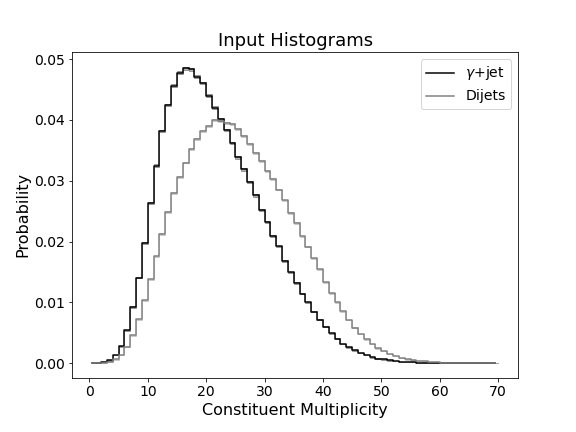}
        \caption{pp input distributions}
        \label{fig:pp-input}
    \end{subfigure}
    \begin{subfigure}{0.23\textwidth}
        \centering
        \includegraphics[width=\textwidth]{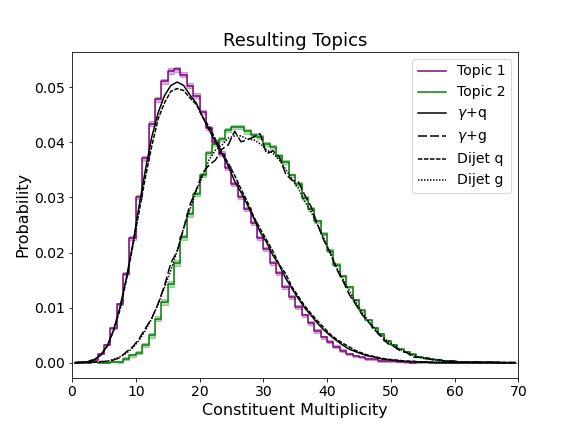}
        \caption{pp topic results}
        \label{fig:pp-topics}
    \end{subfigure}
    \begin{subfigure}{0.23\textwidth}
        \centering
        \includegraphics[width=\textwidth]{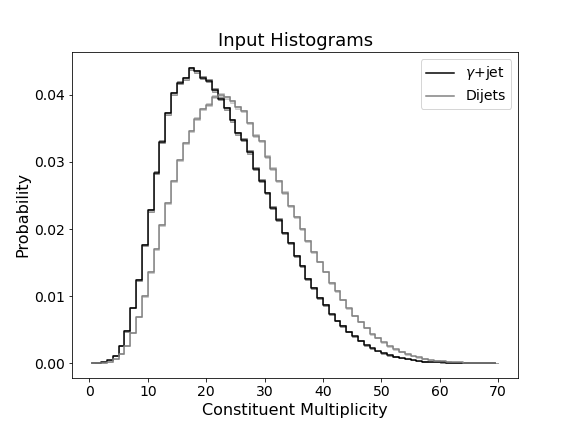}
        \caption{PbPb input distributions}
        \label{fig:hi-input}
    \end{subfigure}
    \begin{subfigure}{0.23\textwidth}
        \centering
        \includegraphics[width=\textwidth]{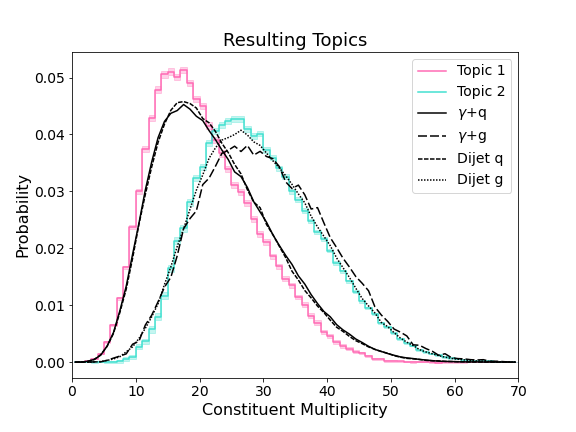}
        \caption{PbPb topic results}
        \label{fig:hi-topics}
    \end{subfigure}
    \caption{\textsc{Pyquen}-generated proton-proton and heavy-ion normalized constituent multiplicity input distributions for $\gamma+$jet and dijets, along with topic extraction results, displaying the resulting topics in comparison to the gluon and quark MC truth labels.}
    \label{fig:inputs}
\end{figure}

In this section, we show the results of the topic modeling algorithm on the \textsc{Pyquen} data. Since previous work~\cite{metodiev-topic-of-jets} has demonstrated that constituent multiplicity approximately satisfies quark-gluon mutual irreducibility, we use this observable as input to the topic modeling. The input distributions and the resulting topics for both the proton-proton and heavy-ion data are shown in Fig.~\ref{fig:inputs}. We also show the MC truth-labeled quark and gluon distributions from $\gamma$+jet and dijet samples. The topic analysis is performed separately for simulated proton-proton and heavy-ion jets, meaning that the topics extracted from the two systems are fully independent.

In general, the extracted topics correspond fairly well to the multiplicity distributions for quark- and gluon-initiated jets as defined from the MC level, with topics 1 and 2 corresponding to quark-like and gluon-like jets, respectively. The agreement between topics and the MC definition is slightly better for proton-proton than for heavy-ion events. The quark-like topics tend to be narrower in both proton-proton and heavy-ion samples, though the effect is exacerbated in heavy-ions. A similar discrepancy was also found in Ref.~\cite{brewer}, based on results from a different Monte Carlo generator, \textsc{Jewel}. We note that the MC-level quark distribution is quite sensitive to the definition of ``quark-initiated'' jets, which is fundamentally ambiguous. A similar discrepancy was also found in Ref.~\cite{brewer} and was discussed in some detail in an Appendix therein. We additionally find a small discrepancy between the MC-level gluon distributions depending on whether they are estimated from the $\gamma$+jet or dijet sample, which could indicate mild sample-dependence of quark and gluon jets or biases due to the MC-level definition of the initiating parton. We note that more jets fail the criterion for MC quark and gluon labeling in heavy-ion compared to proton-proton samples, so the MC-level labeling is more uncertain in this case. On the other hand, the topics are defined on jet samples that include all jets, not just those that are close to one of the leading-order hard matrix elements.

\begin{figure}[htp]
    \centering
    \begin{subfigure}{0.23\textwidth}
        \centering
        \includegraphics[width=\textwidth]{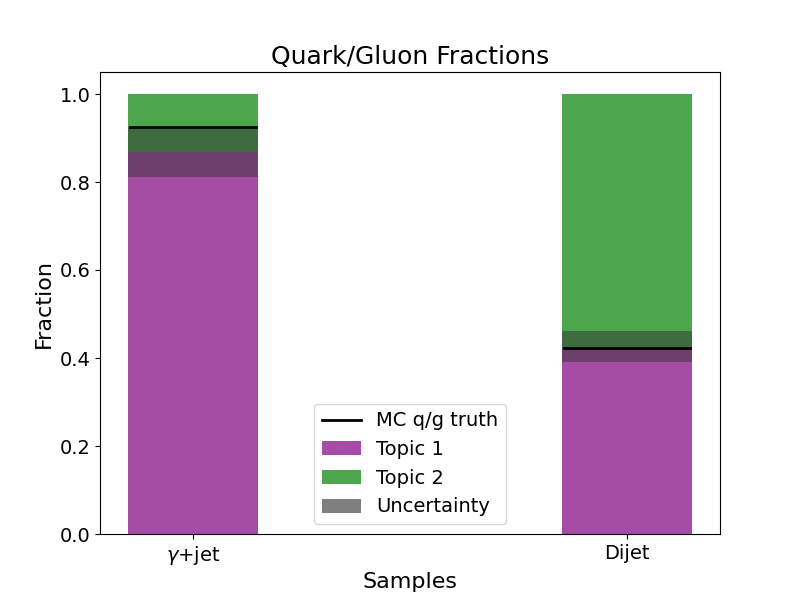}
        \caption{pp fractions}
        \label{fig:pp-qg}
    \end{subfigure}
    \begin{subfigure}{0.23\textwidth}
        \centering
        \includegraphics[width=\textwidth]{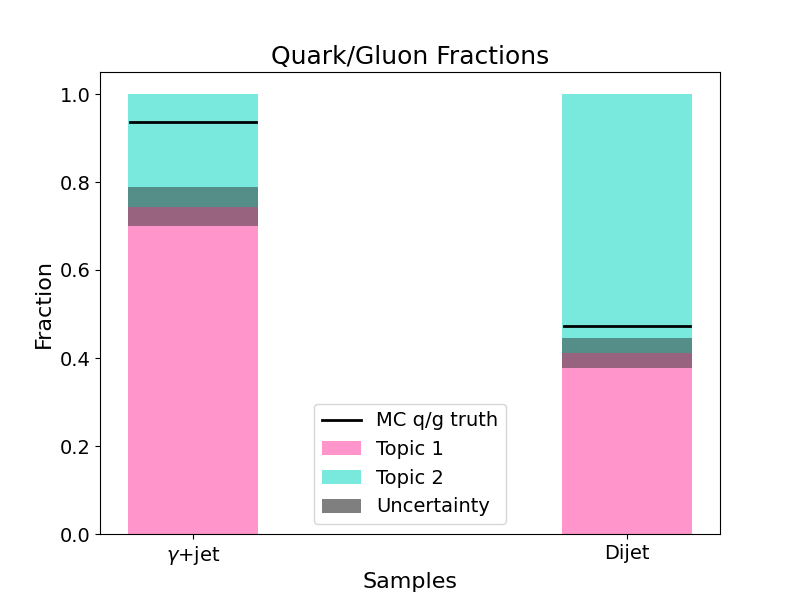}
        \caption{PbPb fractions}
        \label{fig:pbpb-qg}
    \end{subfigure}
    \caption{Quark and gluon fractions in each sample extracted via topic modeling, compared to MC truth quark and gluon fractions.}
    \label{fig:qg-fracs}
\end{figure}

Fig. \ref{fig:qg-fracs} shows the quark and gluon fractions of each sample, extracted from the topic modeling algorithm. The results are compared to the corresponding quark and gluon fractions derived from the MC labels. While the topic and MC fractions are within uncertainties for the proton-proton sample, there are substantial differences in the heavy-ion sample, consistent with differences in the extracted topics. Moreover, the topic modeling algorithm more accurately reproduces the MC-level quark and gluon fractions for dijets than for $\gamma+$jets in both pp and PbPb collisions; this is because the quark-like topic deviates from the MC quark distribution and $\gamma+$jet has a higher quark fraction. These fractions are consistent with those found in \textsc{Jewel} in~\cite{brewer}.

Compared to~\cite{brewer} we also note that the constituent multiplicity distributions themselves are substantially different (primarily due to decays of $\pi^0$ which are allowed in this work). Though this impacts the extracted quark and gluon jet multiplicities, it does not impact the topic modeling algorithm itself as long as those distributions are (approximately) mutually irreducible and the features of quark and gluon jets do not depend on whether they are produced in $\gamma$+jet or dijet events. Excellent agreement between topic modeling results and the MC definition in proton-proton collisions suggests that quark and gluon multiplicities are approximately mutually irreducible and that $\gamma$+jet and dijet samples provide independent fractions of the same underlying quark and gluon jet distributions. Beyond additional ambiguities in the MC labeling, the worse performance of topic modeling in heavy-ion collisions could indicate either lower mutual irreducibility of quark and gluon constituent multiplicity or sample dependence, both of which could potentially arise from interactions with the medium. It is therefore non-trivial that we have found consistent results in \textsc{Pyquen} to those in \textsc{Jewel}, which have completely different descriptions of the medium interaction. This provides hope that topic modeling performance may be comparable in measurements to that seen in these two independent models.

\section{Jet Substructure Extraction}
\label{sec:substructure}
While the data-driven determination of quark- and gluon-like jet fractions in a sample is significant, applying these results to extract quark and gluon jet substructure allows for deeper insight into the modification of quark and gluon jets in the quark-gluon plasma. In this section, we will use the quark and gluon topic fractions extracted from constituent multiplicity distributions in the previous section to extract the quark- and gluon-like distributions for other jet observables.
% demonstrate the application of the topic modeling algorithm to find jet observable distributions corresponding to the topics, and compare these to the MC truth quark and gluon jet observable distributions. 
We will consider the jet shape, jet fragmentation, jet mass, and jet splitting fraction and compare the results to MC-labeled quark and gluon jet distributions. We also utilize these results to determine the modification of quark and gluon jet observables by taking the ratio of the jet observable between heavy-ion and the proton-proton samples, separately for quark- and gluon-like jets.
While the MC-level definition of quark- and gluon-initiated jets is convenient as a qualitative benchmark for the success of this approach, we emphasize that it is both ill-defined and not measurable. The topic modeling procedure, therefore, provides novel access to the separate modification of quark and gluon jet substructure in heavy-ion collisions.

\subsection{Jet Shape}
\begin{figure}[htp]
    \centering
    \begin{subfigure}{0.35\textwidth}
        \centering
        \includegraphics[width=\textwidth]{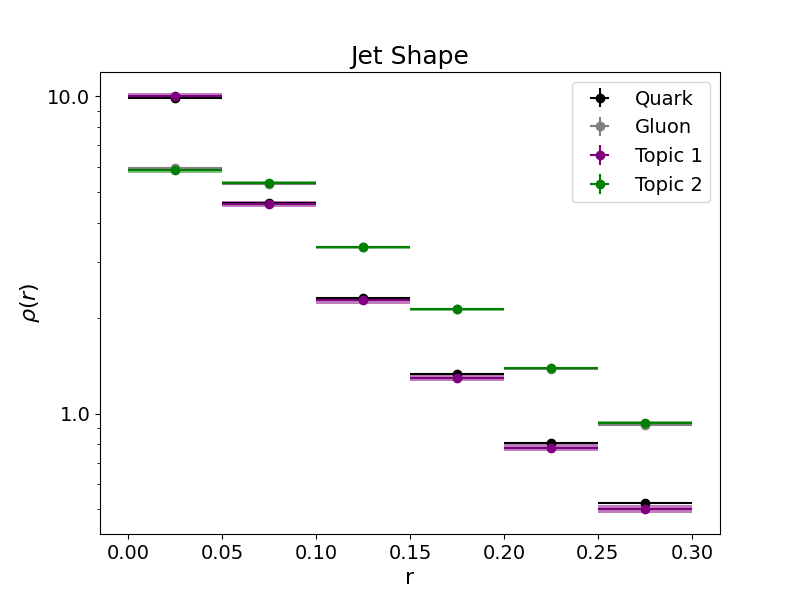}
        \caption{Proton-proton jet shape}
        \label{fig:pp-jet-shape}
    \end{subfigure}
    % \hfill
    \begin{subfigure}{0.35\textwidth}
        \centering
        \includegraphics[width=\textwidth]{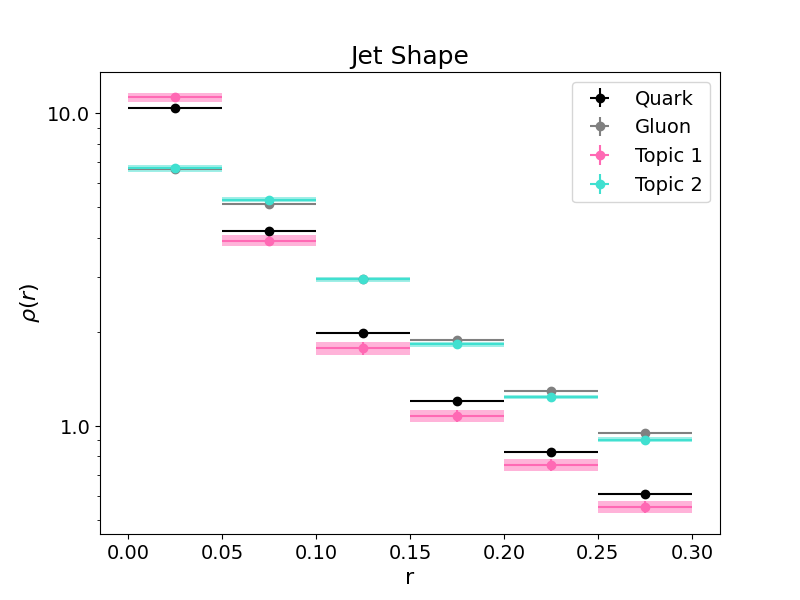}
        \caption{Heavy-ion jet shape}
        \label{fig:hi-jet-shape}
    \end{subfigure}
    % \hfill
    \begin{subfigure}{0.35\textwidth}
        \centering
        \includegraphics[width=\textwidth]{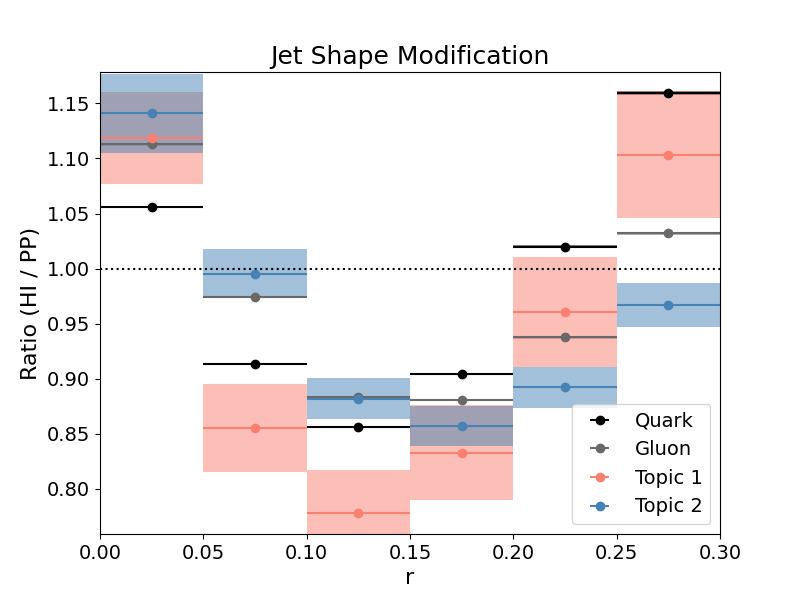}
        \caption{Jet shape modification ratio}
        \label{fig:mod-shape}
    \end{subfigure}
    \caption{Proton-proton (a) and heavy-ion (b) jet shape extraction using topic modeling results from Fig. \ref{fig:inputs}. The jet shape results for the two topics are shown in comparison to the jet shapes constructed from the quark and gluon MC truth labels. The modification of jet shape in the quark-gluon plasma, according to the extracted topics as well as the MC truth labels, is shown as the ratio between heavy-ion and proton-proton jet shape (c).}
    \label{fig:jet-shape}
\end{figure}

Jet shape describes the jet transverse momentum distribution as a function of radial distance from the jet axis, and can be described by the following equation
\begin{equation}
    \rho(r) = \frac{1}{r_b - r_a}\frac{1}{N_\text{jet}} \sum_\text{jets}\frac{\sum_{\text{tracks}\in[r_a, r_b)}p_T^\text{track}}{p_T^\text{jet}}\,.
\end{equation}
Here, $r$ is the radial distance from the jet axis, and $r_a$, and $r_b$ are the inner and outer radii of the given annulus~\cite{jet-shape}. Each annulus corresponds to a bin in the jet shape plot, where $r_a$ is the left edge of the bin and $r_b$ is the right edge of the bin.

In order to obtain the jet shape using our topic modeling results, we can simply perform a linear combination using the extracted $\kappa$ values for each bin in the jet shape:
\begin{equation}
\begin{split}
    \rho_{1}(r) = \frac{\rho_{\gamma+\text{jet}}(r) - \kappa_{AB} \rho_{\text{dijets}}(r)}{1 - \kappa_{AB}},\\
    \rho_{2}(r) = \frac{\rho_{\text{dijets}}(r) - \kappa_{BA} \rho_{\gamma+\text{jet}}(r)}{1 - \kappa_{BA}}\,.
\end{split}
\end{equation}
Here, $\rho_{\gamma+\text{jet}}(r)$ and $\rho_{\text{dijets}}(r)$ are the jet shapes for the $\gamma+\text{jet}$ and $\text{dijets}$, respectively.

In Fig.~\ref{fig:jet-shape} we show the jet shapes extracted from the topic modeling procedure compared to the MC-level distributions of the shape of quark- and gluon-initiated jets. In both proton-proton and heavy-ion collisions, the quark-like topic has a much narrower shape than the gluon-like one, consistent with the MC-level expectation. In proton-proton collisions, the topics are in excellent agreement with the MC definition of quark- and gluon-initiated jets. In heavy-ion collisions, the agreement is qualitative, with the topics being slightly narrower than the MC definition, consistent with the slightly lower multiplicity of topics compared to MC in Fig.~\ref{fig:hi-topics}. The ratio between proton-proton and heavy-ion quark and gluon jet shapes are shown in Fig.~\ref{fig:mod-shape}. The qualitative trend of the topic modification and the MC-level modification are the same, with quark jets having a larger deviation from one at small $r$, presumably due to their steeper jet shape in pp. This feature is enhanced in the topic ratio for quark-like jets since the topic modeling result for the quark-like jet shape is steeper than the MC definition. 

\subsection{Jet Fragmentation Function}
The topic modeling results from Section~\ref{sec:topicresults} can also be used to extract the quark and gluon jet fragmentation function. The jet fragmentation function gives the longitudinal momentum distribution of the tracks inside a jet,
\begin{equation}
    D(\xi) = \frac{1}{N_\text{jet}}\frac{dN_\text{track}}{d\xi}\,.
\end{equation}
Here, $N_\text{jet}$ is the total number of jets, and $N_\text{track}$ is the number of tracks in a jet. $\xi = \ln{(1/z)}$, where $z$ is the longitudinal momentum fraction, is defined as
\begin{equation} 
    z = \frac{p_T\cos{\Delta R}}{p_T^\text{jet}} = \frac{p_T}{p_T^\text{jet}}\cos{\sqrt{(\Delta \eta)^2 + (\Delta \phi)^2}}\,.
\end{equation}
Here, $p_T^\text{jet}$ is the transverse momentum of the jet relative to the beam direction, $p_T$ is the transverse momentum of a charged particle in the jet, and $\Delta \eta$ and $\Delta \phi$ are measures of distance between the particle and E-scheme jet axis in pseudorapidity and azimuth~\cite{jet-frag}.

In the jet fragmentation function, we can also compute each bin of the topics using a linear combination as shown below.
\begin{equation}
\begin{split}
    D_{1}(\xi) = \frac{D_{\gamma+\text{jet}}(\xi) - \kappa_{AB} D_{\text{dijets}}(\xi)}{1 - \kappa_{AB}},\\
    D_{2}(\xi) = \frac{D_{\text{dijets}}(\xi) - \kappa_{BA} D_{\gamma+\text{jet}}(\xi)}{1 - \kappa_{BA}}
\end{split}
\end{equation}
While the jet shape is self-normalized, the jet fragmentation function is normalized by the total number of jets, such that the integral of the histogram over $\xi$ represents the average number of charged particles per jet. Therefore, rather than normalize to get a probability density, we take the direct combination of the per-jet quantities in each bin because we want the output to be a per-jet quantity. By definition,
\begin{equation}
\begin{split}
    D_{\text{dijets}}(\xi) = f_d D_1(\xi) + (1-f_d) D_2(\xi)\\
    D_{\gamma+\text{jet}}(\xi) = f_\gamma D_1(\xi) + (1-f_\gamma) D_2(\xi)
\end{split}
\end{equation}
After we integrate, we arrive at the following set of equations:
\begin{equation}
\begin{split}
    N_{\text{tracks}}^{(\text{dijets})} = f_d N_{\text{tracks}}^{(\text{topic 1})} + (1-f_d) N_{\text{tracks}}^{(\text{topic 2})}\\
    N_{\text{tracks}}^{(\gamma+\text{jet})} = f_\gamma N_{\text{tracks}}^{(\text{topic 1})} + (1-f_\gamma) N_{\text{tracks}}^{(\text{topic 2})}
\end{split}
\end{equation}
which shows that the average number of tracks in dijets (or $\gamma+$jets) is equal to the weighted average of the average number of tracks of each topic. Therefore, rather than include any normalization, we directly apply $\kappa$ to the dijet and $\gamma+$jet jet fragmentation values in order to solve for the jet fragmentation of the topics.

\begin{figure}[htp]
    \centering
    \begin{subfigure}{0.35\textwidth}
        \centering
        \includegraphics[width=\textwidth]{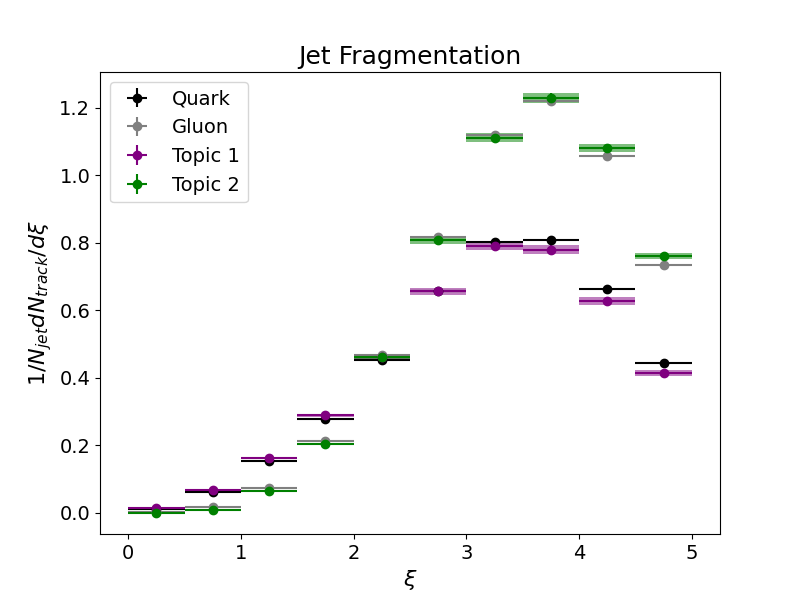}
        \caption{Proton-proton}
        \label{fig:pp-nolog}
    \end{subfigure}
    \begin{subfigure}{0.35\textwidth}
        \centering
        \includegraphics[width=\textwidth]{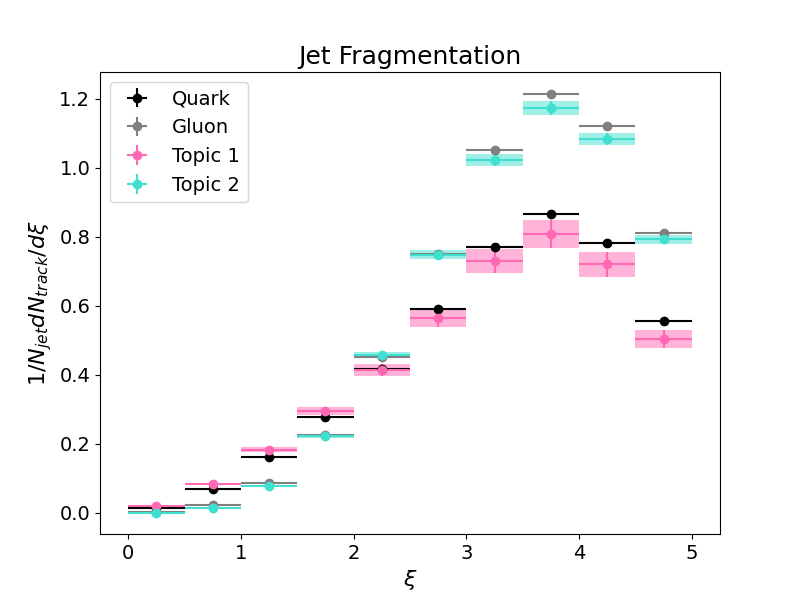}
        \caption{Heavy-ion}
        \label{fig:pbpb-nolog}
    \end{subfigure}
    \begin{subfigure}{0.35\textwidth}
        \centering
        \includegraphics[width=\textwidth]{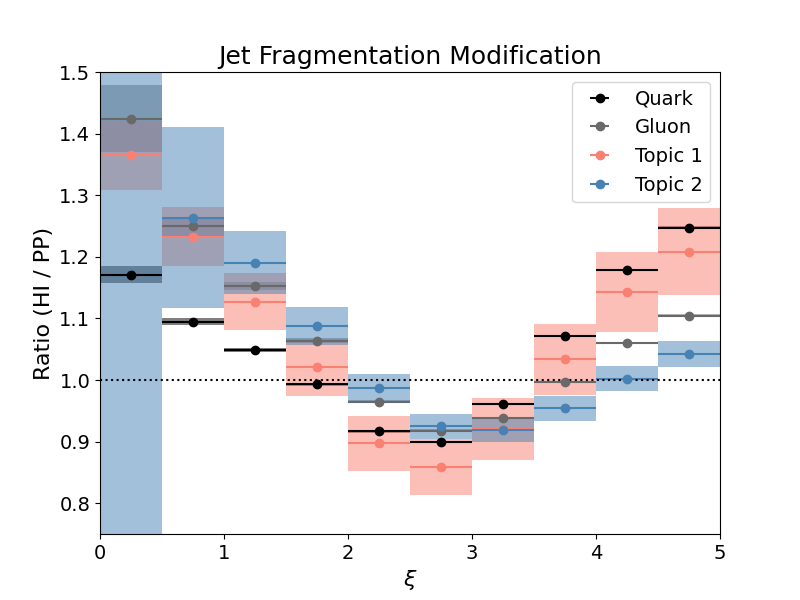}
        \caption{Jet fragmentation modification ratio}
        \label{fig:mod-frag}
    \end{subfigure}
    \caption{Extracting jet fragmentation for proton-proton (a) and heavy-ion (b) collision using topic modeling results from Fig. \ref{fig:inputs}. The jet fragmentation modification, represented by the ratio between PbPb and pp jet fragmentation, is also shown (c) for the topics, as well as quark and gluon.}
    \label{fig:jet-frag}
\end{figure}

Fig.~\ref{fig:jet-frag} shows the extracted jet fragmentation function using topic modeling compared to the fragmentation of MC-defined quark- and gluon-initiated jets. Overall there is a qualitative agreement between the topics and the MC definition over the full range of $\xi$.
Fig.~\ref{fig:mod-frag} shows the ratio between proton-proton and heavy-ion fragmentation functions for quark and gluon topics compared to MC results, which are in qualitative agreement. The quantitative agreement between quark (gluon) ratios at high (low) $\xi$ is apparently accidental since there are differences in the fragmentation in both numerator and denominator.

\subsection{Jet Mass}

In addition to the jet shape and jet fragmentation, which measure the distribution of energy in specified areas of the jet cone, we also demonstrate topic modeling as applied to two additional per-jet substructure observables: jet mass and jet splitting fraction $z_g$.

The jet mass is calculated from the total four-momentum of all the constituents in the jet and is expressed as $m = \sqrt{E^2-|\vec{p}|^2}$, where $E$ is the jet energy and $\vec{p}$ is the momentum of the jet. To extract the jet mass histograms using our topic modeling results, we take a linear combination of the normalized jet mass input histograms, $H_{\gamma+\text{jet}}(m)$ and $H_{\text{dijets}}(m)$, using the extracted $\kappa$ values:
\begin{equation}
\begin{split}
    H_{1}(m) = \frac{H_{\gamma+\text{jet}}(m) - \kappa_{AB} H_{\text{dijets}}(m)}{1 - \kappa_{AB}},\\
    H_{2}(m) = \frac{H_{\text{dijets}}(m) - \kappa_{BA} H_{\gamma+\text{jet}}(m)}{1 - \kappa_{BA}}
\end{split}
\end{equation}

\begin{figure}[htp]
    \centering
    \begin{subfigure}{0.35\textwidth}
        \centering
        \includegraphics[width=\textwidth]{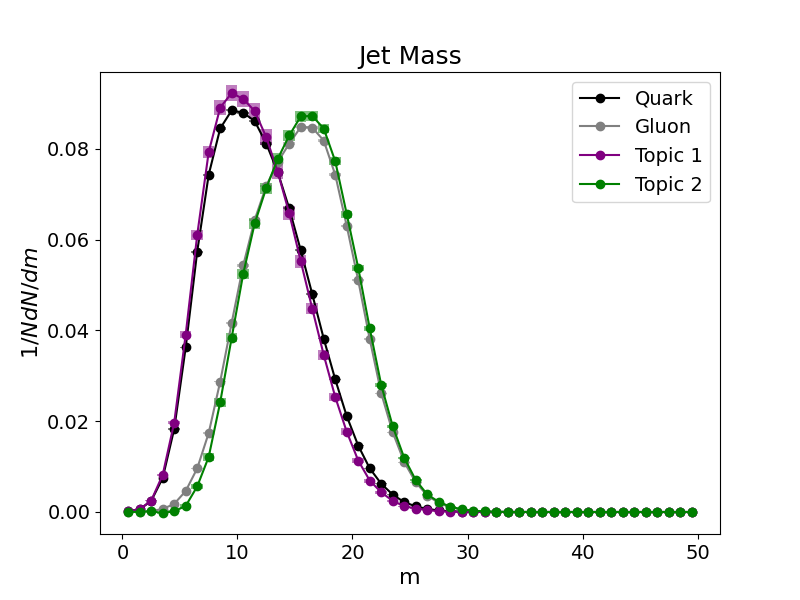}
        \caption{Proton-proton jet mass}
        \label{fig:pp-jet-mass}
    \end{subfigure}
    % \hfill
    \begin{subfigure}{0.35\textwidth}
        \centering
        \includegraphics[width=\textwidth]{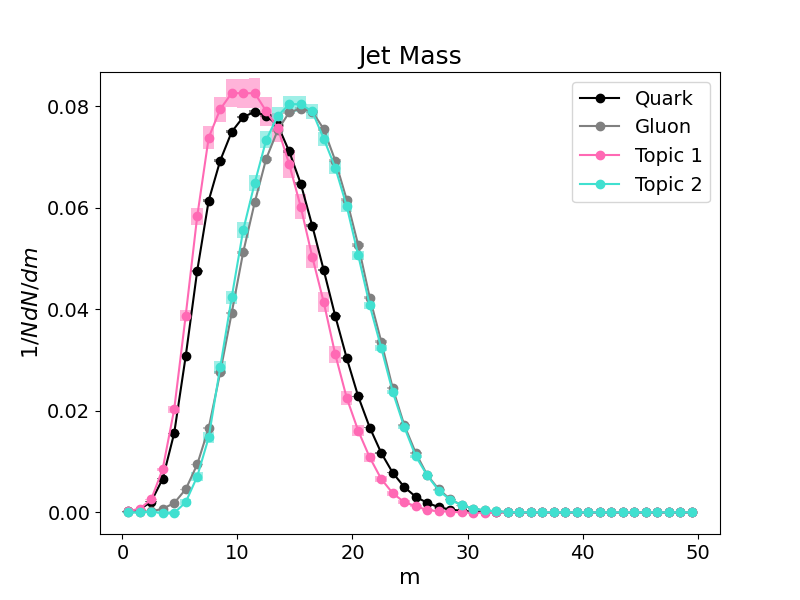}
        \caption{Heavy-ion jet mass}
        \label{fig:hi-jet-mass}
    \end{subfigure}
    % \hfill
    \begin{subfigure}{0.35\textwidth}
        \centering
        \includegraphics[width=\textwidth]{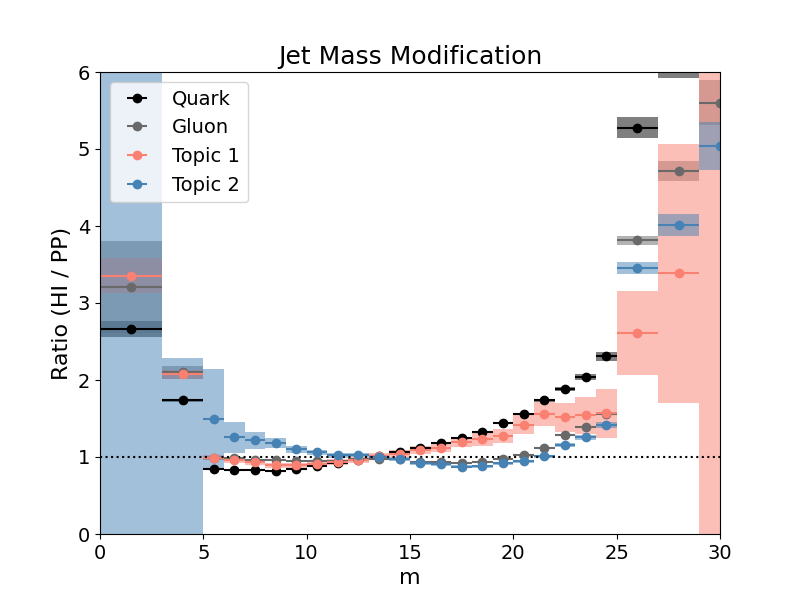}
        \caption{Jet mass modification ratio}
        \label{fig:mod-mass}
    \end{subfigure}
    \caption{Topic modeling results for proton-proton (a) and heavy-ion (b) jet mass, compared to MC truth. Modification of the jet mass in the QGP, as defined by the ratio of the heavy-ion and proton-proton jet mass spectra is shown as well (c).}
    \label{fig:jet-mass}
\end{figure}

The resulting jet mass histograms for quark and gluon topics in pp and PbPb are shown in Fig.~\ref{fig:jet-mass}, along with those for the MC-labelled quark and gluon samples. The ratio between the PbPb and pp jet mass histogram bins is shown in Fig.~\ref{fig:mod-mass}. As before, we find qualitative agreement between the topics and MC-level quark- and gluon-initiated jet distributions, with the quark topic having a slightly lower mass consistent with the lower multiplicity of the quark topic in Fig.~\ref{fig:hi-topics}. Larger deviations from the MC definition in the mass modification ratio at low and high masses are due to the slightly larger deviation of the quark topic in heavy-ions than in proton-proton.

\subsection{Jet Splitting Fraction}
The jet momentum splitting fraction $z_g$ describes the momentum sharing of the first hard splitting inside a jet and is related to the underlying QCD splitting functions. Technically, $z_g$ is the momentum ratio of the leading and subleading subjets for the first splitting in a jet that passes the soft drop condition~\cite{2018-splitting}. 

In order to find the leading and subleading subjets, we use SoftDrop~\cite{softdrop} / mMDT~\cite{mmdt} to decluster the jet's branching history, until the transverse momenta of the subjets fulfill the SoftDrop condition:
\begin{equation}
    \frac{\min{(p_{T,i}, p_{T,j})}}{p_{T,i}+p_{T,j}} > z_{cut}\theta^\beta
\end{equation}
where $\theta$ represents the relative distance in the jet resolution parameter between the two subjets. The settings of SoftDrop used for this analysis were $z_\text{cut} = 0.1$ and $\beta = 0$~\cite{kauder, marzani, 2018-splitting}.

\begin{figure}[htp]
    \centering
    \begin{subfigure}{0.35\textwidth}
        \centering
        \includegraphics[width=\textwidth]{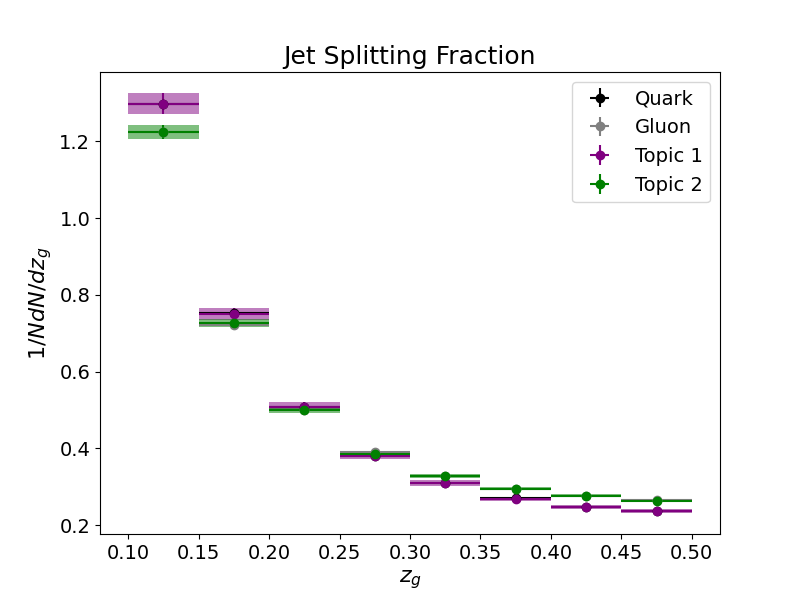}
        \caption{Proton-proton jet splitting fraction}
        \label{fig:pp-jet-zg}
    \end{subfigure}
    % \hfill
    \begin{subfigure}{0.35\textwidth}
        \centering
        \includegraphics[width=\textwidth]{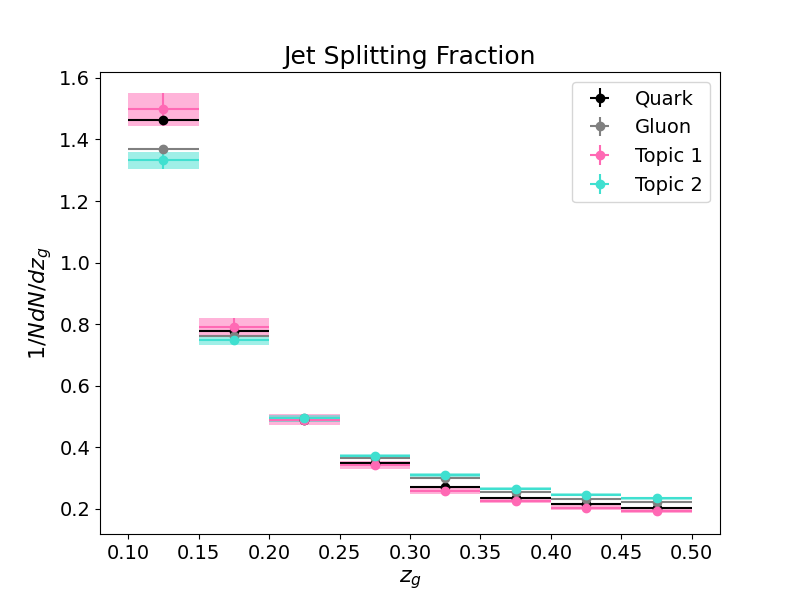}
        \caption{Heavy-ion jet splitting fraction}
        \label{fig:hi-jet-zg}
    \end{subfigure}
    % \hfill
    \begin{subfigure}{0.35\textwidth}
        \centering
        \includegraphics[width=\textwidth]{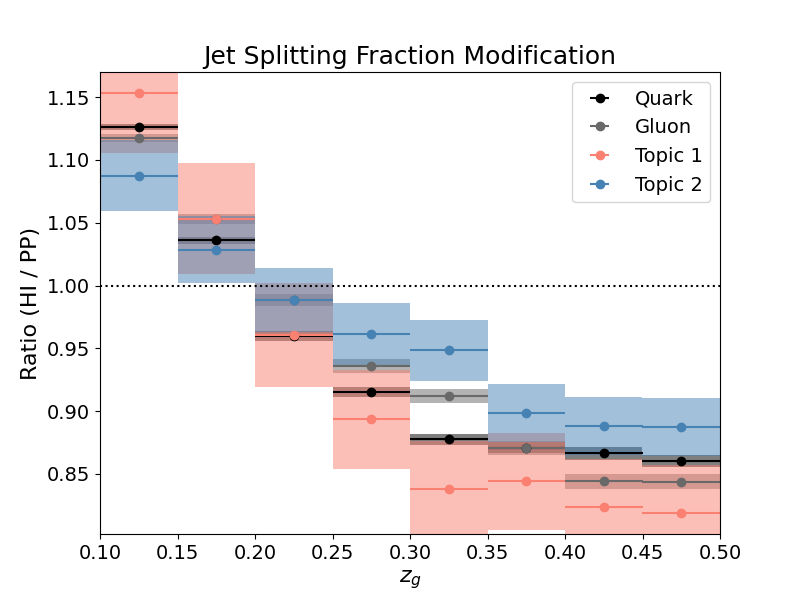}
        \caption{Jet splitting fraction modification ratio}
        \label{fig:mod-zg}
    \end{subfigure}
    \caption{Topic modeling results for proton-proton (a) and heavy-ion (b) jet mass, compared to MC truth. Modification of the jet mass in the QGP, as defined by the ratio of the heavy-ion and proton-proton jet mass spectra is shown as well (c).}
    \label{fig:jet-zg}
\end{figure}

The procedure for extracting the topics is the same as for jet mass. The quark and gluon splitting functions extracted from topics and from the MC definition are shown in Fig.~\ref{fig:jet-zg} for both proton-proton and heavy-ion samples. In this case, the topics agree semi-quantitatively with the MC definition in both proton-proton and heavy-ion samples, and in their ratio Fig.~\ref{fig:mod-zg}. The better agreement between the topics and MC definition in this observable compared to others shown in this Section may be due to the comparatively weak dependence on the quark and gluon fractions.

\section{Simulating Thermal Background}
\label{sec:thermal}
In heavy-ion collisions, particles from the jets are accompanied by a large background of particles from the quark-gluon plasma that adds additional particles in the jet radius. Even with effective background subtraction, this still contributes to non-negligible smearing of the properties of jets in experimental data. These effects contribute in addition to the \textsc{Pyquen}-generated PbPb distributions shown in this paper, which only include the jets themselves without effects from the background. Since the topic separation ultimately depends on multiplicity distributions which may be especially sensitive to these effects, in this section we estimate the effectiveness of the topic modeling in the presence of such smearing.

\begin{figure}[htp]
    \centering
    \begin{subfigure}{0.23\textwidth}
        \centering
        \includegraphics[width=\textwidth]{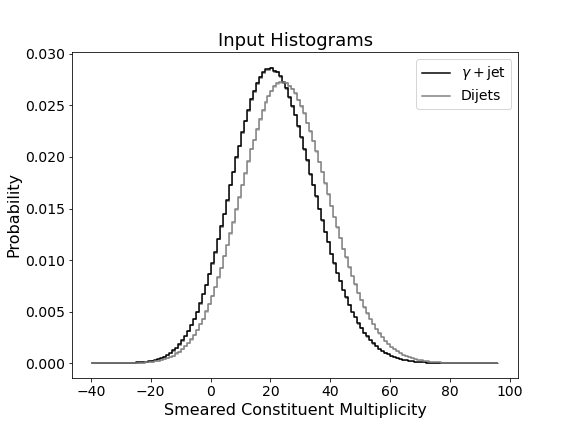}
        \caption{Proton-proton smeared constituent multiplicity}
        \label{fig:pp-smeared}
    \end{subfigure}
    \begin{subfigure}{0.23\textwidth}
        \centering
        \includegraphics[width=\textwidth]{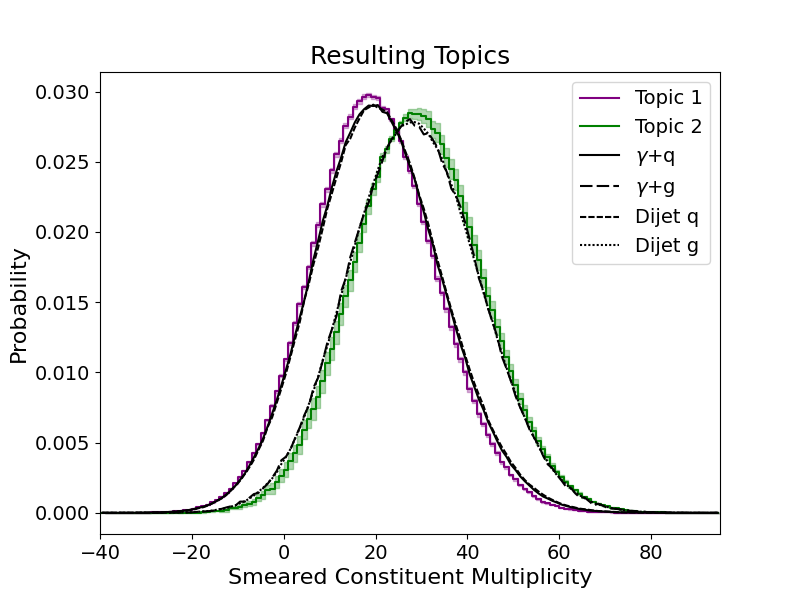}
        \caption{Topic modeling results (pp)}
        \label{fig:pp-smeared-results}
    \end{subfigure}
    \begin{subfigure}{0.23\textwidth}
        \centering
        \includegraphics[width=\textwidth]{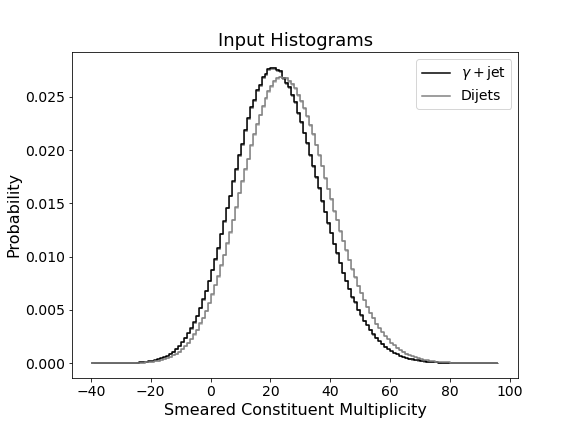}
        \caption{Heavy-ion smeared constituent multiplicity}
        \label{fig:pbpb-smeared}
    \end{subfigure}
    \begin{subfigure}{0.23\textwidth}
        \centering
        \includegraphics[width=\textwidth]{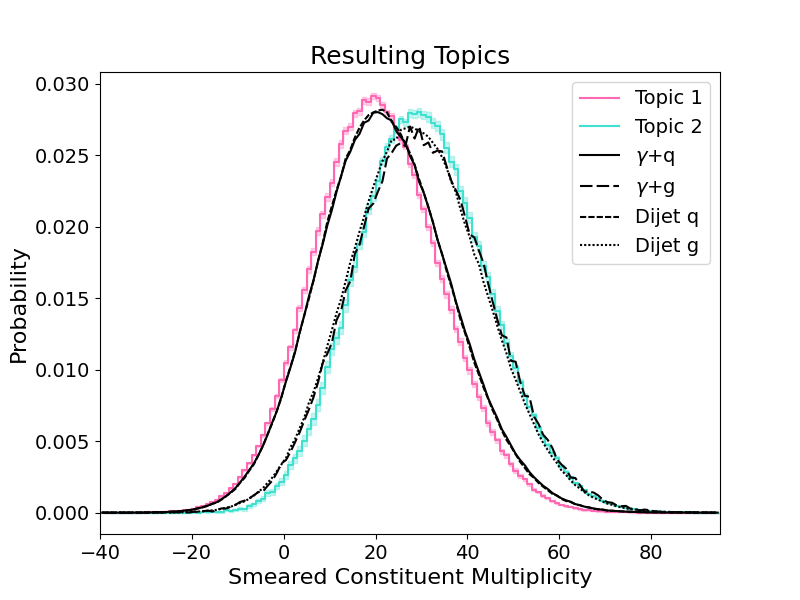}
        \caption{Topic modeling results (PbPb)}
        \label{fig:pbpb-smeared-results}
    \end{subfigure}
    \caption{Smeared constituent multiplicity inputs (left) with corresponding topic separation results (right)}
    \label{fig:mult-smeared}
\end{figure}

We consider smearing both the proton-proton and heavy-ion multiplicity distributions with this background and performing the topic separation on those smeared distributions. In proton-proton collisions, the smearing would be done for example through embedding in minimum-bias heavy-ion events. We use $d N/d\eta\, d\phi \sim 1600/(2\pi)$~\cite{CMS:2011aqh} to estimate that the number of particles from the background inside of a cone of $R=0.4$ is $\sim dN/d\eta\,d\phi (\pi R^2) \sim 11^2$. Assuming that background subtraction can eliminate the average, we estimate the expected fluctuations of particles within the jet cone as a Gaussian distribution $\mathcal{N}(0, 11)$. 
The resulting smeared histograms and smeared MC-labeled quark and gluon distributions are shown in Fig. \ref{fig:mult-smeared}. To extract reducibility factors we fit these distributions with 2 skew-normal and 2 normal underlying distributions, rather than with 4 skew-normal distributions as in the original method.

With smearing, the input multiplicity distributions for $\gamma+$jets and dijets are less distinguishable. However, the topic separation is still capable of performing well, which demonstrates the robustness of the algorithm to changes in the input distributions.

\begin{figure}[htp]
    \centering
    \begin{subfigure}{0.23\textwidth}
        \centering
        \includegraphics[width=\textwidth]{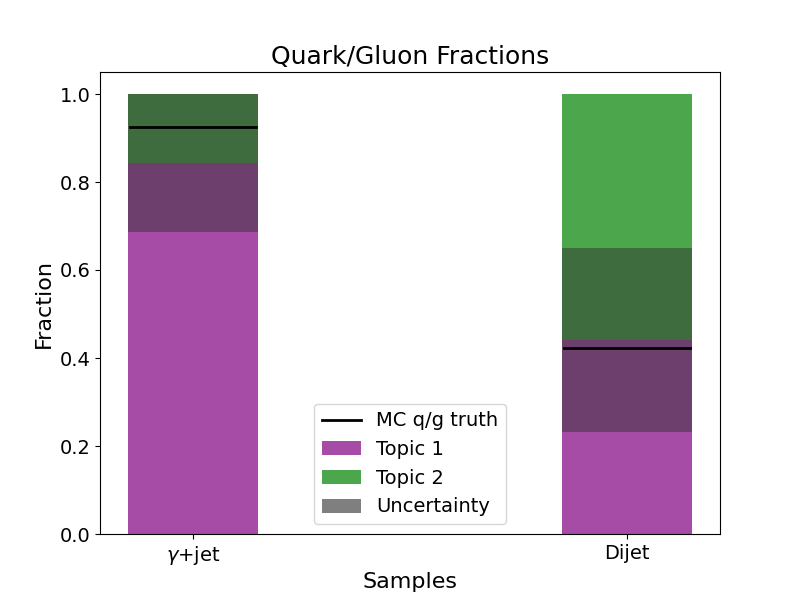}
        \caption{Jet fractions in pp sample}
        \label{fig:pp-smeared-fracs}
    \end{subfigure}
    \begin{subfigure}{0.23\textwidth}
        \centering
        \includegraphics[width=\textwidth]{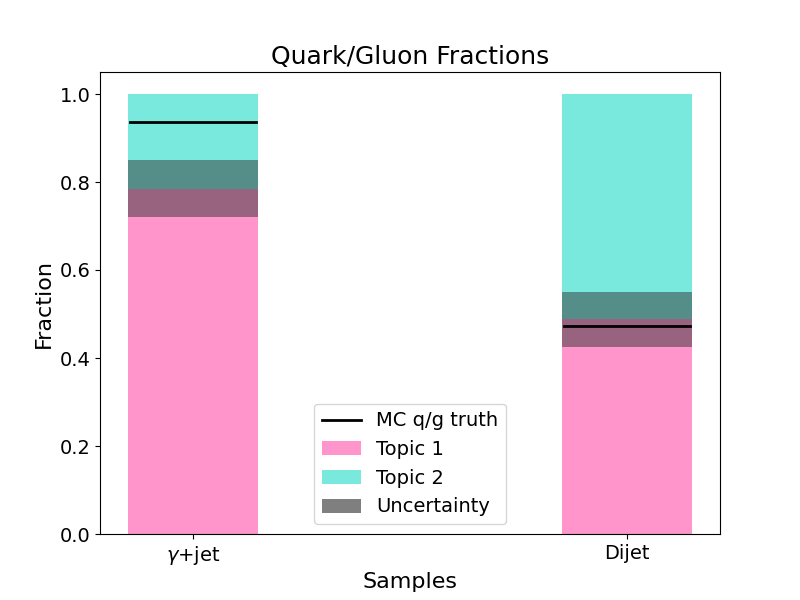}
        \caption{Jet fractions in PbPb sample}
        \label{fig:pbpb-smeared-fracs}
    \end{subfigure}
    \caption{Topic separation jet fractions displayed against MC truth quark and gluon truth values}
    \label{fig:frac-smeared}
\end{figure}

The quark-like fractions for $\gamma+$jets and dijets in each sample are shown in Fig. \ref{fig:frac-smeared}. The smeared quark and gluon fraction values are similar to those of the unsmeared dataset in Fig. \ref{fig:qg-fracs}. The uncertainty is much larger when background fluctuations are included because the absolute tails of the distributions (where the topics should be extracted) are at the tail of both the ``true'' multiplicity distributions and the Gaussian smearing distribution. Nonetheless, the consistency of the results indicates that the topic separation is resistant to constituent multiplicity fluctuations in the jet cone due to large backgrounds.

The resulting quark/gluon substructure extraction using the smeared pp and heavy-ion datasets are in Appendix~\ref{app:substructures}. As in the main text, we find qualitative agreement between the extracted jet substructure and MC-level quark- and gluon-initiated jet distributions. The uncertainty on the pp results is significantly larger, while the increase in uncertainty in PbPb is not as drastic. Despite the precision of the topic separation deteriorating, the accuracy of calculated substructure values remains similar to the unsmeared dataset. This is also reflected in the modification plots comparing the ratio of the smeared heavy-ion substructure values to the unsmeared proton-proton substructure values.

\section{Discussion and Conclusion}
\label{sec:conclusions}

In summary, our results from \textsc{Pyquen}-generated Monte Carlo samples corroborate previous proof-of-concept studies performed using \textsc{Jewel}, and demonstrate that a fully data-driven technique can potentially be used to extract separate quark and gluon jet distributions from experimental samples, without additional knowledge or templates. We extend the previous study by demonstrating that the resulting fractions can be used to extract jet substructure observables for quark- and gluon-like jets from $\gamma$+jet and dijet substructure measurements. As a proof-of-principle, we showed results for the jet shape, fragmentation function, jet mass, and splitting function separately for quark- and gluon-like jet topics. We additionally found that these results are robust to smearing the multiplicity distributions used to extract the topics by a large background as in heavy-ion collisions. These results suggest potential for an experimental determination of quark and gluon jet spectra and their substructure using this technique.

The resulting topics and their modification are in good qualitative agreement with the MC-level definition for quark- and gluon-initiated jets. There are quantitative discrepancies, which could result from ambiguities in the MC-level definition of quark and gluon jets. We define the MC labels to only include jets that have a quark or gluon outgoing matrix element within the jet radius, which does not include all jets in the sample. This implies that the input $\gamma+$jet and dijet samples, from which we extract topics, are not pure mixtures of Monte Carlo-labelled quark and gluon jets. Discrepancies could also arise from minor violations of the assumptions of the topic modeling algorithm. For example, if constituent multiplicities of quark and gluon jets are not fully mutually irreducible in heavy-ion collisions, different observables (for example, derived from machine learning) may be required to yield better results~\cite{komiske} and could be an interesting avenue for future work.

We have found that the substructure observables obtained from the topics provide a robust estimate of the quark and gluon jet substructure modification in the quark-gluon plasma. This provides a powerful technique to study quark and gluon jet modification in the quark-gluon plasma. If measured, quark and gluon jet substructure would provide strong additional constraints on the theory and modeling of jet interactions with the quark-gluon plasma. 

\section*{Code Availability}
The code for the topic modeling algorithm and the subsequent substructure observable extraction can be found at \url{https://github.com/kying18/jet-topics}.

\vspace{20pt}

\section*{Acknowledgements}

We are grateful to Jesse Thaler and Nima Zardoshti for their valuable discussions on this topic and for their detailed comments on the manuscript. This work has been supported by the Department of Energy, Office of Science, under Grant No. DE-SC0011088 (to Y.Y., Y.C., and Y.L.).

\appendix
\section{Smeared Substructures and Modification Plots} \label{app:substructures}

\begin{figure}[htp]
    \centering
    \begin{subfigure}{0.35\textwidth}
        \centering
        \includegraphics[width=\textwidth]{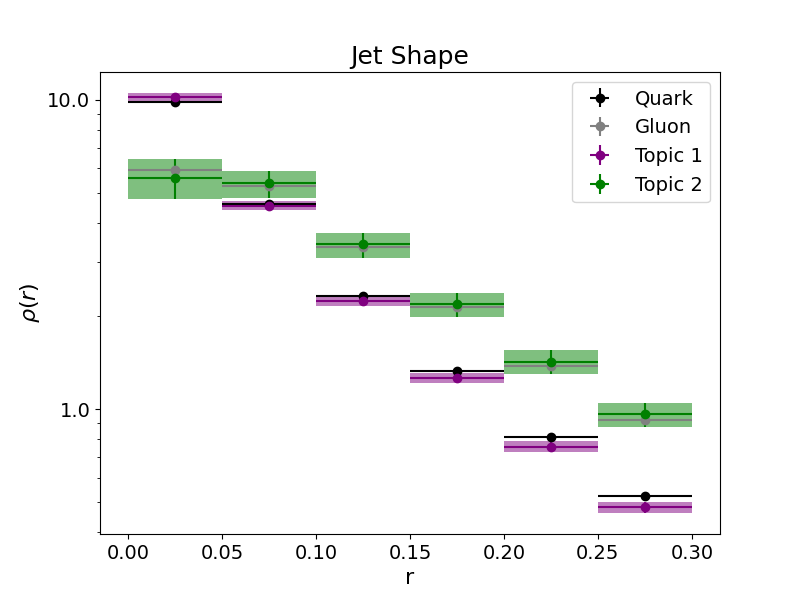}
        \caption{pp jet shape}
        % \label{fig:}
    \end{subfigure}
    % \hfill
    \begin{subfigure}{0.35\textwidth}
        \centering
        \includegraphics[width=\textwidth]{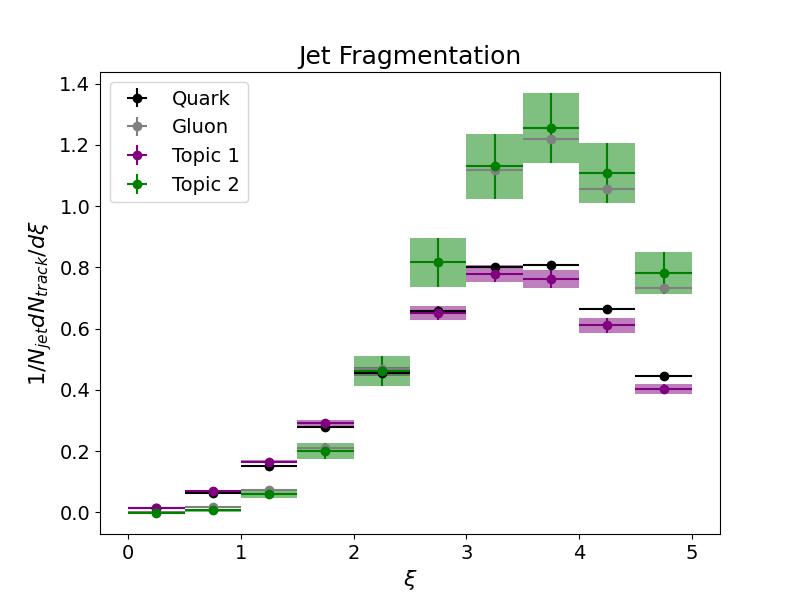}
        \caption{pp jet fragmentation}
        \label{}
    \end{subfigure}
    % \hfill
    \begin{subfigure}{0.35\textwidth}
        \centering
        \includegraphics[width=\textwidth]{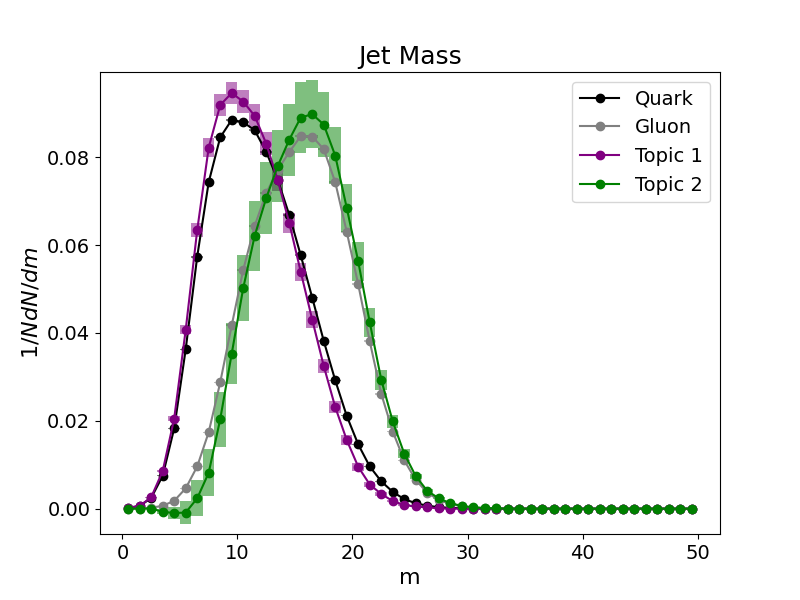}
        \caption{pp jet mass}
        \label{}
    \end{subfigure}
    \begin{subfigure}{0.35\textwidth}
        \centering
        \includegraphics[width=\textwidth]{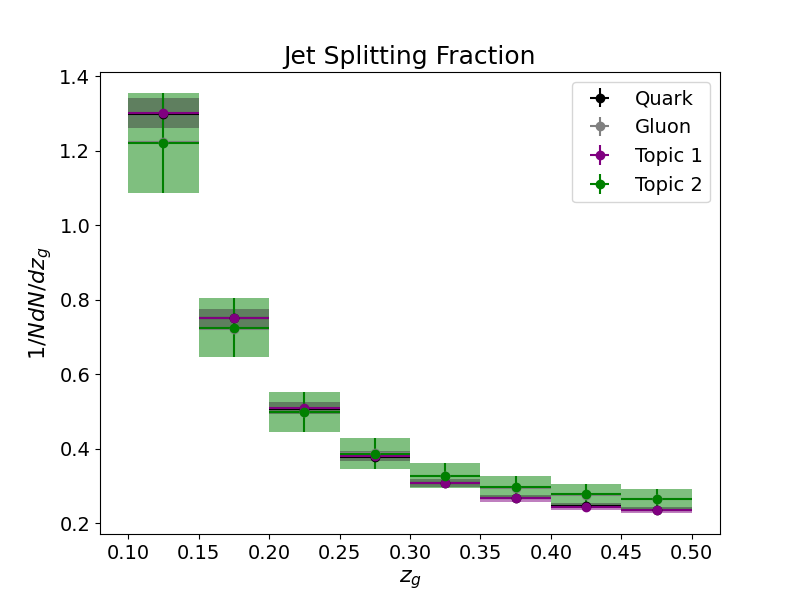}
        \caption{pp jet splitting fraction}
        \label{}
    \end{subfigure}
    \caption{Proton-proton quark/gluon jet substructures extracted from smeared data}
    \label{fig:smeared-pp-substructures}
\end{figure}

\begin{figure}[htp]
    \centering
    \begin{subfigure}{0.35\textwidth}
        \centering
        \includegraphics[width=\textwidth]{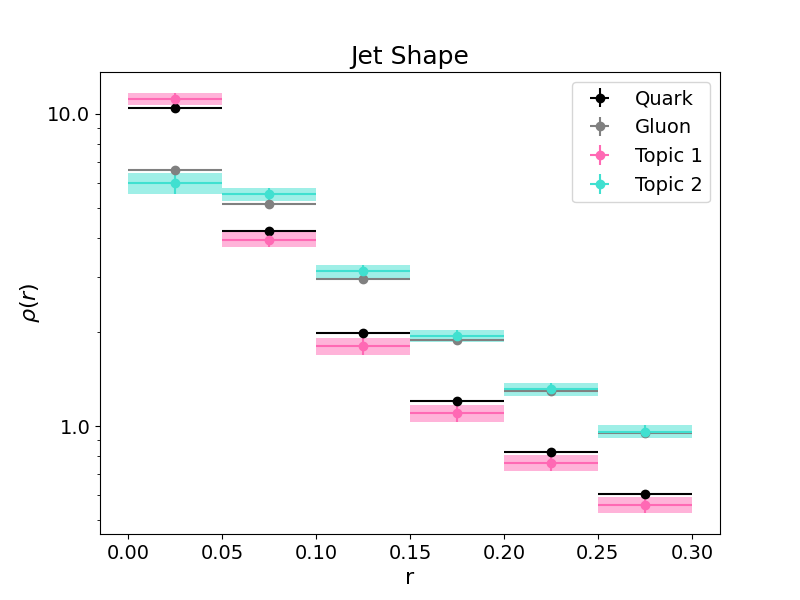}
        \caption{PbPb jet shape}
        % \label{fig:}
    \end{subfigure}
    % \hfill
    \begin{subfigure}{0.35\textwidth}
        \centering
        \includegraphics[width=\textwidth]{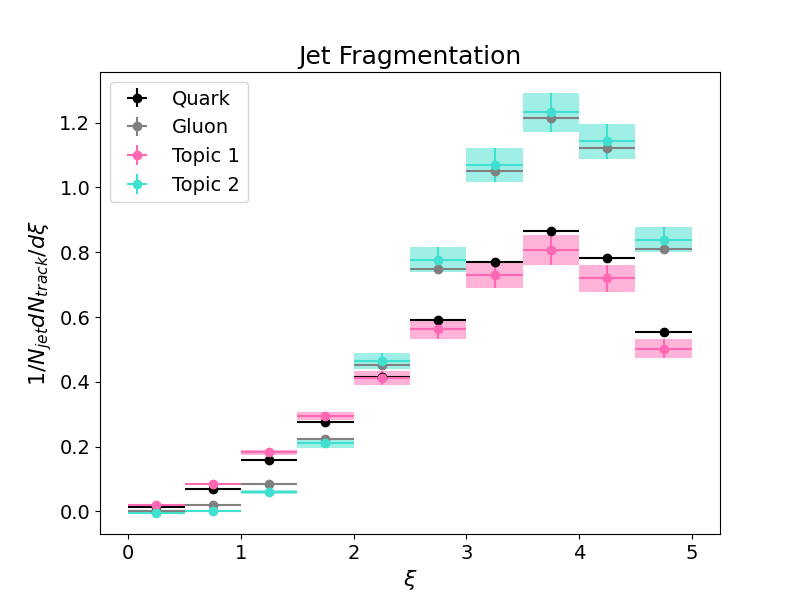}
        \caption{PbPb jet fragmentation}
        \label{}
    \end{subfigure}
    % \hfill
    \begin{subfigure}{0.35\textwidth}
        \centering
        \includegraphics[width=\textwidth]{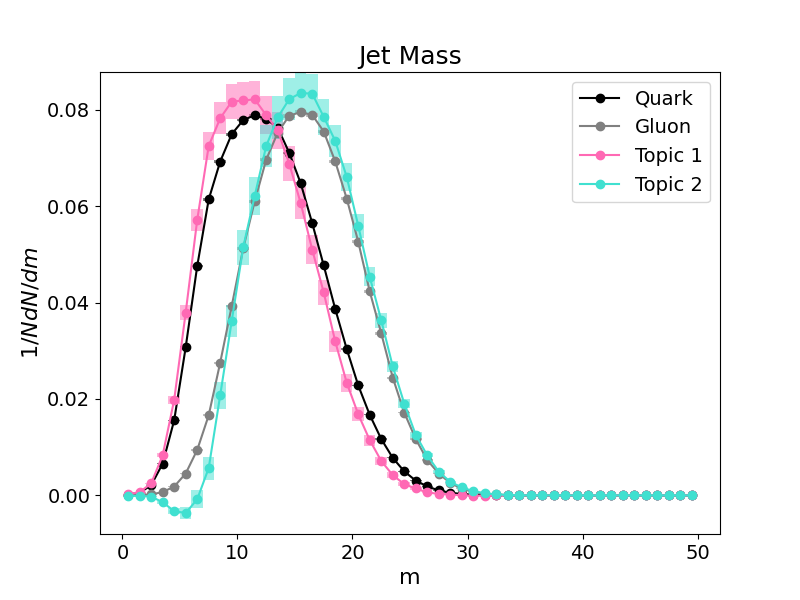}
        \caption{PbPb jet mass}
        \label{}
    \end{subfigure}
    \begin{subfigure}{0.35\textwidth}
        \centering
        \includegraphics[width=\textwidth]{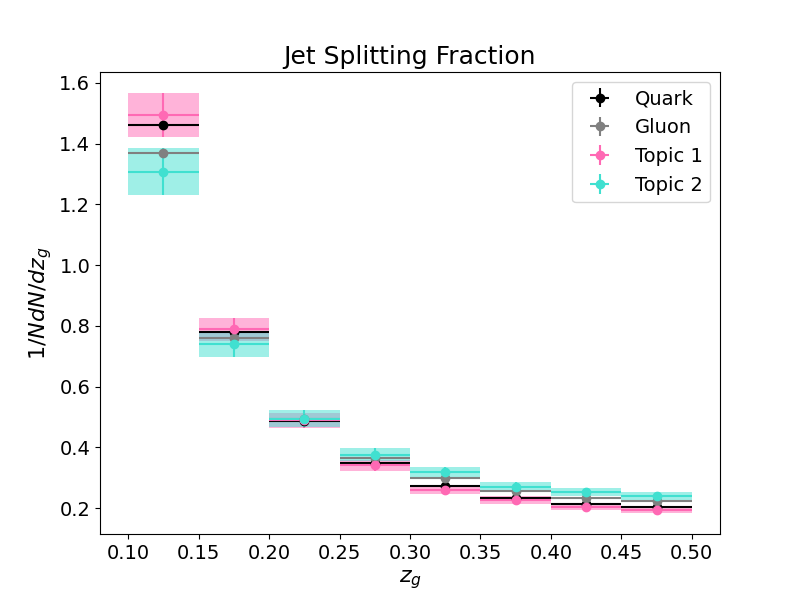}
        \caption{PbPb jet splitting fraction}
        \label{}
    \end{subfigure}
    \caption{Heavy-ion quark/gluon jet substructures extracted from smeared data}
    \label{fig:smeared-pbpb-substructures}
\end{figure}

The quark and gluon jet substructures extracted from the smeared constituent multiplicity are shown in Fig. \ref{fig:smeared-pp-substructures} (pp) and Fig. \ref{fig:smeared-pbpb-substructures} (PbPb). While the extracted substructures are quite similar between the smeared data and unsmeared data, the uncertainty is notably larger. In addition, the topic 2 jet mass calculation falls slightly negative in both the pp and PbPb samples, which is not physically attainable. This is due to high $\kappa$ values on the left tail of the extraction. % TODO need to add more here but not quite sure what to say?

\begin{figure}[htp]
    \centering
    \begin{subfigure}{0.35\textwidth}
        \centering
        \includegraphics[width=\textwidth]{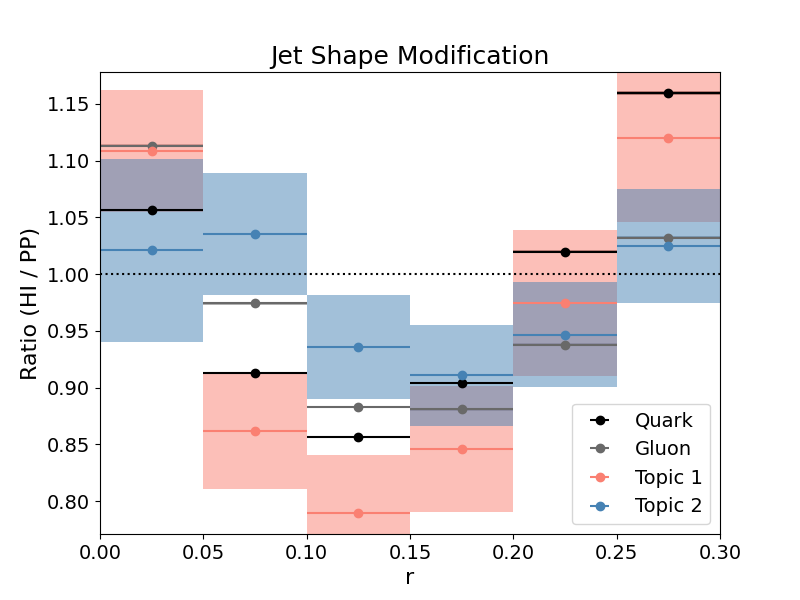}
        \caption{Jet shape modification}
        % \label{fig:}
    \end{subfigure}
    % \hfill
    \begin{subfigure}{0.35\textwidth}
        \centering
        \includegraphics[width=\textwidth]{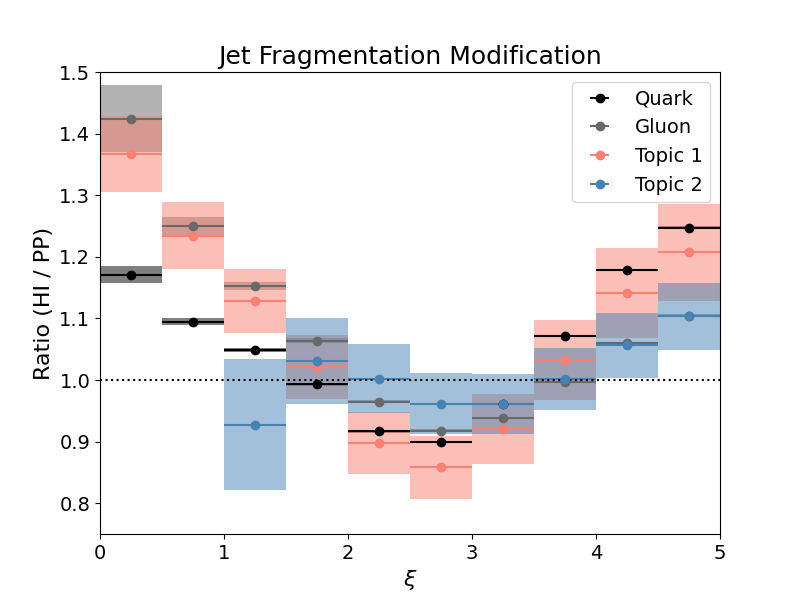}
        \caption{Jet fragmentation modification}
        \label{}
    \end{subfigure}
    % \hfill
    \begin{subfigure}{0.35\textwidth}
        \centering
        \includegraphics[width=\textwidth]{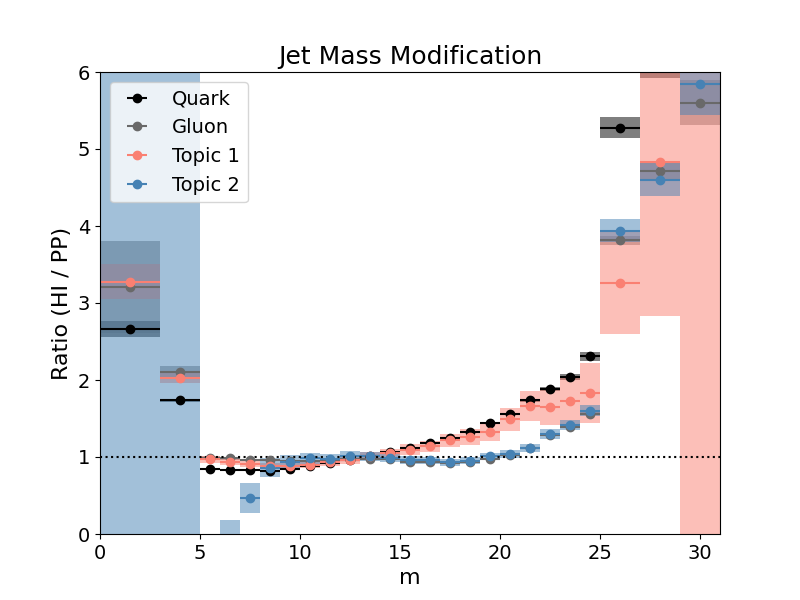}
        \caption{Jet mass modification}
        \label{}
    \end{subfigure}
    \begin{subfigure}{0.35\textwidth}
        \centering
        \includegraphics[width=\textwidth]{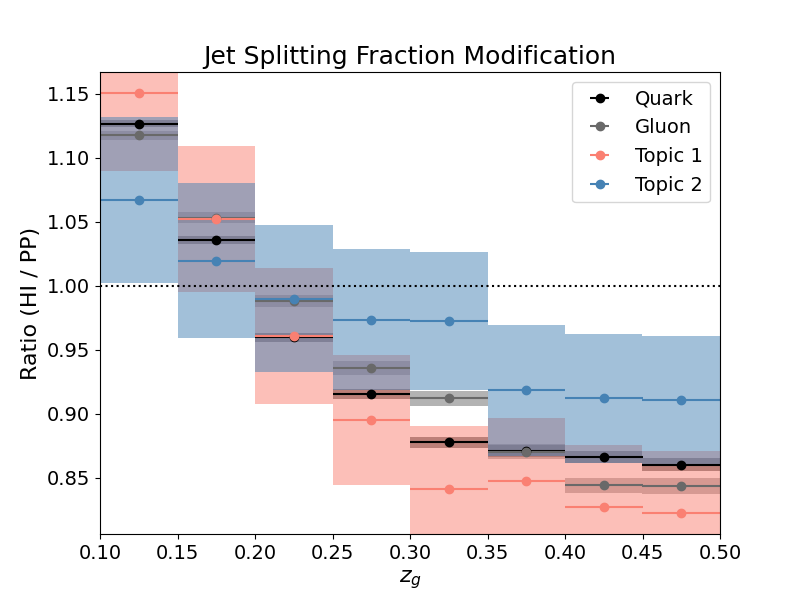}
        \caption{Jet splitting fraction modification}
        \label{}
    \end{subfigure}
    \caption{QGP modification comparing smeared heavy-ion substructure results to unsmeared proton-proton substructure results}
    \label{fig:mod-plots-smeared}
\end{figure}

We also display the modifications in the QGP in Fig. \ref{fig:mod-plots-smeared}, which are determined using the ratio between the smeared heavy-ion jet substructure and the unsmeared proton-proton jet substructure. 

\bibliographystyle{unsrt}
\bibliography{apssamp}% Produces the bibliography via BibTeX.

\end{document}